# RADIATIVE RECOMBINATION SPECTRA OF HEAVILY p-TYPE δ-DOPED GaAs/AlAs MQWs

J. KUNDROTAS[a,b*], A. ČERŠKUS[a], G. VALUŠIS[a],
M. LACHAB[c], S. P. KHANNA[c], P. HARRISON[c] AND E. H. LINFIELD[c]

[a] Semiconductor Physics Institute, A. Goštauto 11, LT-01108 Vilnius, Lithuania

[b] Gediminas' Technical University of Vilnius, Saulėtekio 11, LT-10223 Vilnius, Lithuania

[c] Institute of Microwave and Photonics, University of Leeds, Leeds LS2 9JT, United Kingdom

We present a study of the photoluminescence (PL) properties of heavily Be δ-doped GaAs/AlAs multiple quantum wells measured at room and liquid nitrogen temperatures. Possible mechanisms for photocarriers recombination are discussed, with a particular focus on the peculiarities of the excitonic and free carriers-acceptors PL emissions occurring below and above the Mott metal-insulator transition. Moreover, based on a simple theoretical model, it is found that the critical impurities concentration to observe the Mott transition in the MQW samples exhibiting 15 nm wells width and 5 nm-thick barrier layers is $\sim 3 \times 10^{12}$ cm$^{-2}$.

## 1. Introduction

The investigation of the impurities in two-dimensional (2D) semiconductor structures is thus of fundamental interest and could lead to novel device applications such as quantum well (QW) infrared emitters and photodetectors. During the growth of doped epitaxial QW layers, the impurities could be either homogenously distributed along the growth direction or localized within a single atomic layer, the so called δ-doping. In the case of highly δ-doped bulk semiconductors or quantum wells, the impurity-related PL spectra suffer drastic changes. Indeed, with increasing the impurity concentrations, the interacting doping species form new states in the valence band, which lead to the creation of a 2D carriers gas and occurrence of the Mott transition [1-7]. In this article, we report the results of photoluminescence (PL) experiments carried out on highly Be δ-doped GaAs/AlAs multiple QWs at liquid nitrogen and room temperatures.

---

* corresponding author; e-mail: kundrot@pfi.lt

## 2. Samples growth and experimental results

The multiple quantum wells (MQWs) structures were grown on semi-insulating GaAs substrates using molecular beam epitaxy. The samples consist of 60 GaAs wells grown to a thickness, $L_W$, of 15 nm, and separated by 5 nm thick AlAs barriers. The QWs were δ-doped at the center with Be acceptors: Four doping levels, $N_{Be}$, were investigated, namely: $2.7\times10^{11}$ cm$^{-2}$ (sample #1), $2.7\times10^{12}$ cm$^{-2}$ (#2), $2.7\times10^{13}$ cm$^{-2}$ (#3) and $5.3\times10^{13}$ cm$^{-2}$ (#4).

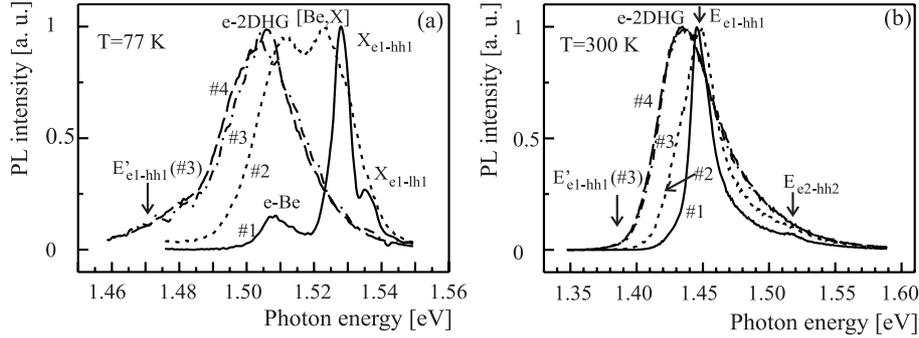

**Fig. 1.** PL spectra of the Be δ-doped GaAs/AlAs MQWs recorded (a) at 77 K and (b) at room temperature. Arrows labelled $E_{e1-hh1}$ and $E_{e2-hh2}$ indicate the calculated transition energies for first heavy-hole to first electron and second heavy-hole to second electron, respectively. $e$-2DHG refer to the optical transitions associated with the 2D hole gas. $E'_{e1-hh1}$(#3) is the energy difference between sublevels in sample #3.

A continuous wave argon-ion laser was used to excite the PL. The optical spectra measured at room and liquid nitrogen temperatures are shown in Fig. 1. First, we discuss the data obtained for the weakly doped sample, #1. At cryogenic temperature, the PL spectrum reveals the presence of a series of well-resolved peaks (see Fig. 1a). The main PL emission bands denoted $X_{e1-hh1}$ and $X_{e1-lh1}$ are ascribed to the heavy- and light-hole excitonic transitions, respectively. The lower energy transitions, labelled $e$-Be, results from the recombination of free electrons with Be acceptors [6]. As for the line [Be, $X$], it originates from the emission of excitons bound to acceptors.

The spectra of medium and highly doped samples are more complicated. The PL spectrum recorded at 77 K for the sample #2 clearly depicts the transitions $e$-Be and [Be, $X$] while the emissions $X_{e1\text{-}hh1}$ and $X_{e1\text{-}lh1}$ appear rather as shoulders. Spectra at room temperature are similar to weakly doped sample #1. The spectra of highly doped samples #3 and #4 exhibit very different PL spectra. As one can see, changes are observed in fine structure and, secondly, spectra are shifted to lower energy side. It will be shown that the radiative recombination is related with formation 2D hole gas in these samples.

### 3. Model and discussion

The model of noninteracting impurities may only satisfactorily explain the optical properties of the weakly δ-doped MQWs. However, with increasing the dopants concentration, the single-impurity theory becomes no longer valid as a result of the significant overlap between the electronic wavefunctions of adjacent impurities. With increase of dopant concentration starts the formation of the impurity band and tail of subband edge. At very high doping levels the Mott transition then occurs [5]. In δ-doped QWs, whose impurity concentrations are above the metallic limit, the impurities start forming the so-called V-shaped potential wells with a new 2D subband structure [1, 2].

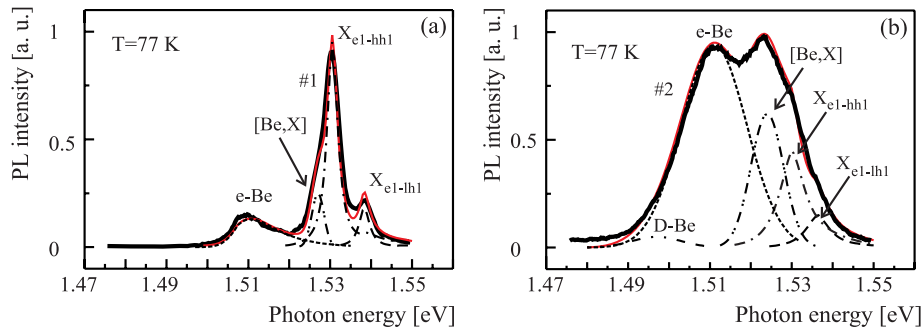

**Fig. 2.** PL spectra of (a) the weakly doped (#1) and (b) medium doped (#2) GaAs/AlAs MQW samples measured at 77 K. Solid lines represent the experimental results while all other lines illustrate theoretical approximations of the various recombination mechanisms of photocarriers highlighted in Fig. 1.

To analyze the optical data of the moderately δ-doped GaAs/AlAs MQWs (sample #2), we assume that the quantum wells are nondegenerate and their impurity concentrations are close to the Mott transition. The PL spectra feature a few kinks, which are consistent with the energies of excitonic transitions observed in the weakly doped sample. The shapes of the PL emission lines for the different transitions were determined using the fractional dimensional space approach [6]. The results are shown in Fig. 2.

In the high-density limit, the holes are confined to the sheet of ionized impurities and form a 2D hole gas with a 2D subband structure. The critical impurity concentration, $N_I$, for the insulator-metal transition and formation of the subband structure in δ-doped structures can be estimated from the relation: $N_I^{1/2} a_B \approx 0.31$ [4, 7], where $a_B$ is the Bohr radius of the impurity. Based on a simple spherical symmetry model for acceptor species in GaAs/AlAs QWs, and assuming an activation energy $E_{Be}$=32 meV, that is $a_{Be}$=1.8 nm, the critical concentration is found to be $N_{Be}$=2.85×10$^{12}$ cm$^{-2}$. This result means that in heavily doped samples forms 2D hole gas, and the PL spectra extent region from lowest energy levels up to Fermi energy level. The energy levels were calculated from the self-consistent solutions of Poisson and Schrödinger equations. The simulation results also showed that for the structure #3, the optical transition $E'_{e1\text{-}hh1}$(#3) occurs at 1.47 eV, at 77 K. This value agrees well with the experiments as shown in Fig. 1.

## Acknowledgements

The work was partly supported by the Lithuanian State Science and Studies Foundation under contract C-07004.

## References

[1] *Delta-doping of semiconductors*, Ed. E.F. Schubert (Cambridge University Press, 1996).

[2] J.J. Harris, R. Murray, and C.T. Foxon, *Semicond. Sci. Technol.* **8**, 31 (1993).

[3] A. John Peter, and K. Navaneethakrishnan, *Physica E* **16**, 223 (2003).

[4] J. Kundrotas, A. Čerškus, S. Ašmontas, A. Johannessen, G. Valušis, B. Sherliker, M.P. Halsall, P. Harrison, and M.J. Steer, *Lithuanian J. Phys.* **45**, 201 (2005).

[5] N.F. Mott, *Metal-insulator transitions*, (Taylor & Francis Ltd, London 1974).

[6] J. Kundrotas, A. Čerškus, S. Ašmontas, G. Valušis, B. Sherliker, M.P. Halsall, M.J. Steer, E. Johannessen, and P. Harrison, *Phys. Rev. B* **72**, 235322 (2005).

[7] J. Kortus, and J. Monecke, *Phys. Rev. B* **49**, 17216 (1994).